\documentclass[letterpaper, 10 pt, conference]{ieeeconf}  % Comment this line out if you need a4paper

\IEEEoverridecommandlockouts                              % This command is only needed if 
\usepackage{graphics} % for pdf, bitmapped graphics files
\usepackage{epsfig} % for postscript graphics files
\usepackage{mathptmx} % assumes new font selection scheme installed
\usepackage{times} % assumes new font selection scheme installed
\usepackage{amsmath} % assumes amsmath package installed
\usepackage{amssymb}  % assumes amsmath package installed
\usepackage{hyperref}
\usepackage{colortbl}
\usepackage{comment}
\usepackage{subcaption}
\usepackage{xcolor}

\title{\LARGE \bf
5G-Enabled Smart Prosthetic Hand: Connectivity Analysis and Assessment
}

\author{Ozan Karaali$^{1}$, Hossam Farag$^{1}$, Strahinja Do\v sen$^{2}$, \v Cedomir Stefanovi\' c$^{1}$ \\
$^{1}$Department of Electronic Systems, Aalborg University, Denmark\\
$^{2}$Department of Health
Science and Technology, Aalborg University, Denmark\\
Email: \{ozank,hmf,cs\}@es.aau.dk, sdosen@hst.aau.dk
}

\begin{document}

\maketitle
\thispagestyle{empty}
\pagestyle{empty}

%%%%%%%%%%%%%%%%%%%%%%%%%%%%%%%%%%%%%%%%%%%%%%%%%%%%%%%%%%%%%%%%%%%%
\begin{abstract}
In this paper, we demonstrate a proof-of-concept implementation of a framework for the development of edge-connected prosthetic systems.
The framework is composed of a bionic hand equipped with a camera and connected to a Jetson device that establishes a wireless connection to the edge server, processing the received video stream and feeding back the inferred information about the environment. 
The hand-edge server connection is obtained either through a direct 5G link, where the edge server also functions as a 5G base station, or through a WiFi link.
We evaluate the latency of closing the control loop in the system, showing that, in a realistic usage scenario, the connectivity and computation delays combined are well below 125 ms, which falls into the natural control range.
To the best of our knowledge, this is the first analysis
showcasing the feasibility of a 5G-enabled prosthetic system. 

\end{abstract}

%%%%%%%%%%%%%%%%%%%%%%%%%%%%%%%%%%%%%%%%%%%%%%%%%%%%%%%%%%%%%%%%%%%%%%%%%%%%%%%%
\section{INTRODUCTION}
Robotic hand prostheses can be used to restore motor functions lost due to amputation. Modern devices are mechatronically advanced, but effective interfacing to these systems, allowing users to exploit all the available functions easily and reliably, is still missing. In the conventional approach to prosthesis control, users activate their muscles to generate command signals (electromyography, EMG) for the prosthesis. The signals can be mapped directly to the target degrees of freedom in the case of simple systems, or they can be interpreted using machine learning (pattern classification and regression) to recognize many movements, which are then replicated by an advanced multifunctional and/or dexterous prosthesis \cite{marinelli_active_2023}. In these control schemes, the user has full control of the system and needs to generate explicit commands for each and every action, which can be cognitively taxing, especially when controlling advanced systems with many functions. 

An approach to address this challenge is to make prostheses smarter by enhancing them with additional sensing and processing, endowing them with context awareness to perform some functions automatically, thus decreasing the cognitive burden of control \cite{guo_toward_2023}. This approach represents the application of shared control, well-known in robotics and automation, to the context of prosthetics. A typical semi-autonomous prosthesis prototype uses an RGB camera or a depth sensor placed on the prosthesis (or integrated into the hand) to recognize the object the user wants to grasp. Computer vision methods, such as deep learning or point cloud analysis, are then used to estimate the object properties (e.g., shape, size, and orientation). Fusing this information with the data from other sensors (e.g., inertial measurement units or prosthesis encoders), the prosthesis controller automatically decides the wrist angle and hand configuration suitable for grasping. 

As demonstrated in~\cite{mouchoux_artificial_2021}, semi-autonomous prostheses can improve performance while decreasing physical and mental efforts; however, they also require complex processing (e.g., computer vision and sensor fusion). Such processing is too demanding to run locally on the system itself, and most prototypes presented in the literature are demonstrated by offloading the computation to a dedicated lab computer. While this showcases the approach, this scheme is not suitable for clinical applications, where users need to operate the device in their daily lives outside the laboratory environment. However, the emergence of cloud and edge computing allows for bridging this barrier, and more generally, opens up unprecedented possibilities in assistive robotics, enabling previously unattainable capabilities with traditional local processing approaches. The critical enabler of such advanced assistive systems is connectivity, which has to be capable of supporting real-time and potentially data-intensive information exchanges between the remote controller and the assistive device.

\begin{figure*}[t]
\centering
\includegraphics[width=0.58\linewidth]{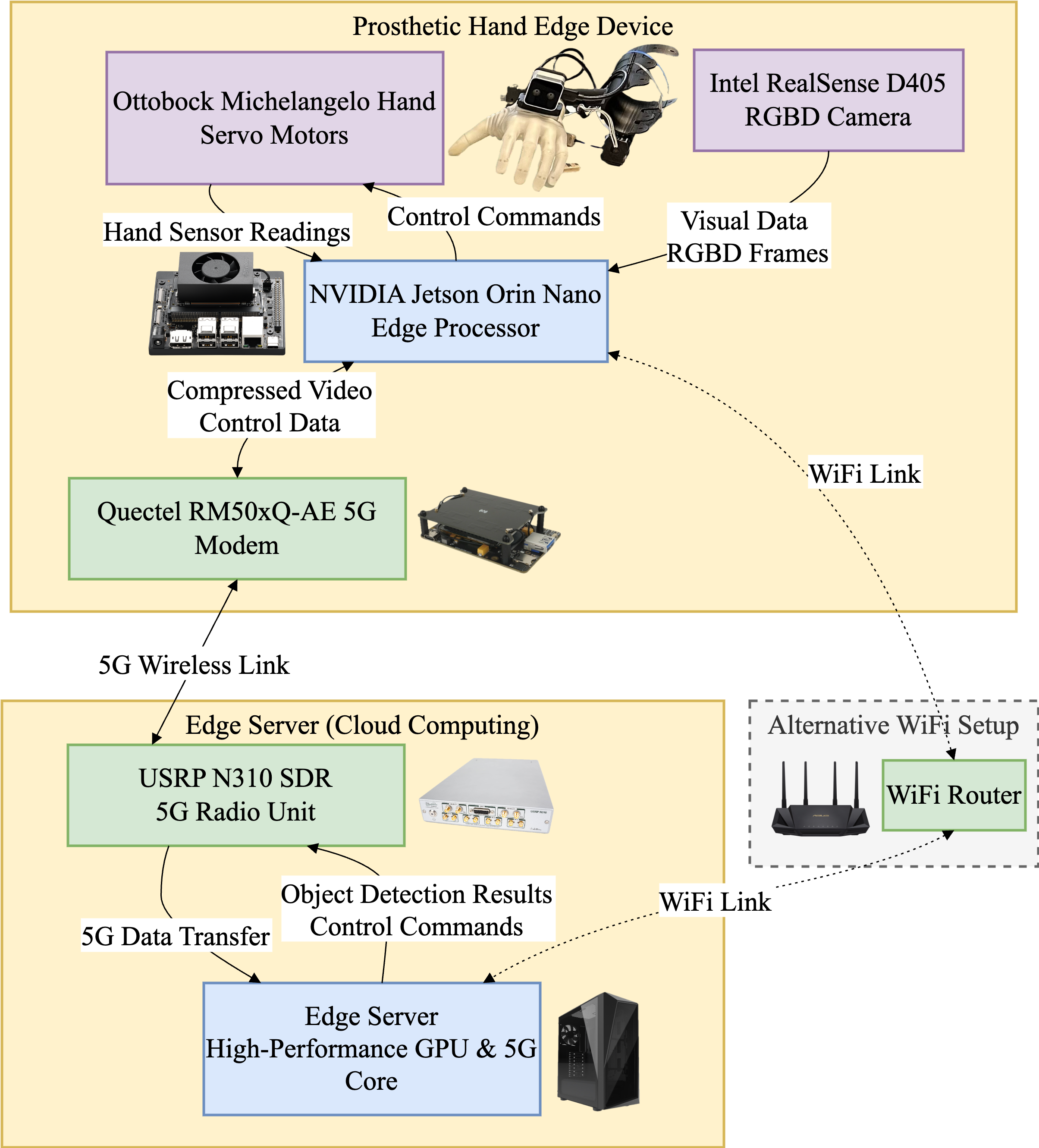} 
\caption{System architecture of the edge-connected prosthetic hand.
Unidirectional arrows indicate the data flow in the system, whereas bidirectional arrows represent two-way communication between the prosthetic device and the edge server. }
\label{fig:systemcomponents}
\end{figure*}

Recent advancements in wireless systems led researchers to shift their focus beyond solely local processing. Shatilov et al. implemented a 3D-printed prosthetic hand that employs a consumer-grade multichannel EMG amplifier and a mobile phone companion application, allowing computation offloading to a cloud server~\cite{shatilov_using_2019}. This smartphone-mediated approach uses a multi-hop method, in which the prosthetic hands connect over a Bluetooth link to the smartphone, which then utilizes a cellular link to the cloud.
Although cost-effective, this solution features delays exceeding 300 ms.

In this respect, it was pointed out that optimal prosthesis control necessitates response times from 100 to 125 milliseconds for natural movement~\cite{farrell_optimal_2007}. To this end, modern 5G networks, with their ultra-reliable low-latency communication and enhanced mobile broadband capabilities~\cite{3gpp.22.261.v15}, can potentially meet stringent timing requirements while enabling direct cloud/edge integration. A direct 5G connectivity between the prosthetic system and the cloud can markedly reduce latencies and enhance bandwidth for data transmission, establishing a foundation for real-time computer vision-based environmental analysis and adaptive control.

This study presents a proof-of-concept framework to implement a 5G-connected prosthesis, demonstrating the feasibility of direct cloud/edge integration for prosthetic control. To the best of our knowledge, this is the first work showcasing such an advanced assistive system that so far has only been envisioned on a conceptual level~\cite{chiariotti_future_2024}. The implemented system consists of an Ottobock Michelangelo prosthetic hand, an Intel RealSense D405 RGB-D camera, a Quectel RM502Q-AE 5G modem, and an edge computing server, demonstrating a realistic implementation that may enable advanced control schemes while maintaining natural response times. The study also assesses the feasibility of online communication between the prosthesis and the edge by measuring end-to-end latencies in a data flow representative of current state-of-the-art semi-automatic prostheses prototypes. Importantly, we show that the offloading of computationally intensive tasks to the cloud and the integration of edge computing achieves end-to-end delays well below 125 ms, which falls within the natural control range~\cite{farrell_optimal_2007}. 

The rest of the paper is organized as follows: We first provide an overview of the edge-connected prosthesis framework, detailing its system architecture, processing pipeline, control flow, communication protocol stack, and software optimizations for real-time performance. Subsequently, we perform the evaluation of the system, including the experimental setup, measurement methodology, and performance analysis of the connectivity and processing pipeline in terms of latency. We conclude the paper with a discussion of the contributions and limitations of the implemented prototype, including an outline of further work.

\begin{comment}

\end{comment}

\section{Overview of the Edge-Connected Prosthetic Framework}

We developed the first framework that can support the design and assessment of 5G-connected smart prostheses prototypes. The starting point is a semi-automatically controlled prosthesis that relies on computer vision for object detection, recognition, and analysis, enabling the automatic selection of a grasping strategy. The prosthesis is equipped with a camera that, when triggered by the user, takes a snapshot of the scene in front of the prosthesis. The snapshot is analyzed using machine learning to determine the object identity and based on this, automatically preshapes the hand, similar to the approach used in \cite{ghazaei_deep_2017}. In the present work, the aforementioned pipeline was implemented using computational offloading to the edge via 5G connectivity, as explained below.

\subsection{System Architecture}

The architecture of the implemented system is depicted in Fig.~\ref{fig:systemcomponents}. At the user's end, the system consists of an Ottobock Michelangelo prosthetic hand, which is capable of performing a range of natural grasping patterns, including palmar grip and lateral grip. To enable environment-aware decision-making, a compact RGB-D camera (Intel RealSense D405) is mounted on the prosthetic hand\footnote{
Future designs will consider integrating the camera directly into the prosthesis.}, providing both visual and depth perception. This camera continuously captures the surrounding environment, allowing the system to analyze objects in real-time. The system integrates an NVIDIA Jetson Orin Nano, which handles local data pre-processing for efficient transmission over the 5G link. The local system is connected to a remote, high-performance edge server via a 5G network to perform computationally intensive tasks such as real-time object detection and control decision-making.
The edge server, being part of the cloud infrastructure, is in the proximity of the network access point to minimize latency, following the distributed computing model where processing resources are deployed near end users.
The 5G network is established using a Quectel RM50xQ-AE modem attached to the Jetson device, which establishes a direct, high-bandwidth link to a 5G base station, whose radio unit is implemented using a USRP N310 software-defined radio (SDR) operating in the n77 frequency band (3.8–4.2 GHz). At the same time, the higher layer functions are performed by the srsRAN software running on the edge server.

The system also features an alternative connectivity option based on a 5 GHz WiFi link.
In this configuration, an ASUS RT-AX58U router is connected to the edge server via Ethernet, while the Jetson device communicates with the router over WiFi. This dual setup facilitates an assessment of the latency and throughput differences between the emerging 5G connectivity solution and a conventional Wi-Fi-based one. The two configurations are both practically relevant, as the WiFi connection can be used when the prosthesis is indoors (e.g., home or work), while the direct 5G link will be employed when there is no access to WiFi (e.g., outdoors). 

\begin{figure}[t]
\centering
\includegraphics[width=1\linewidth]{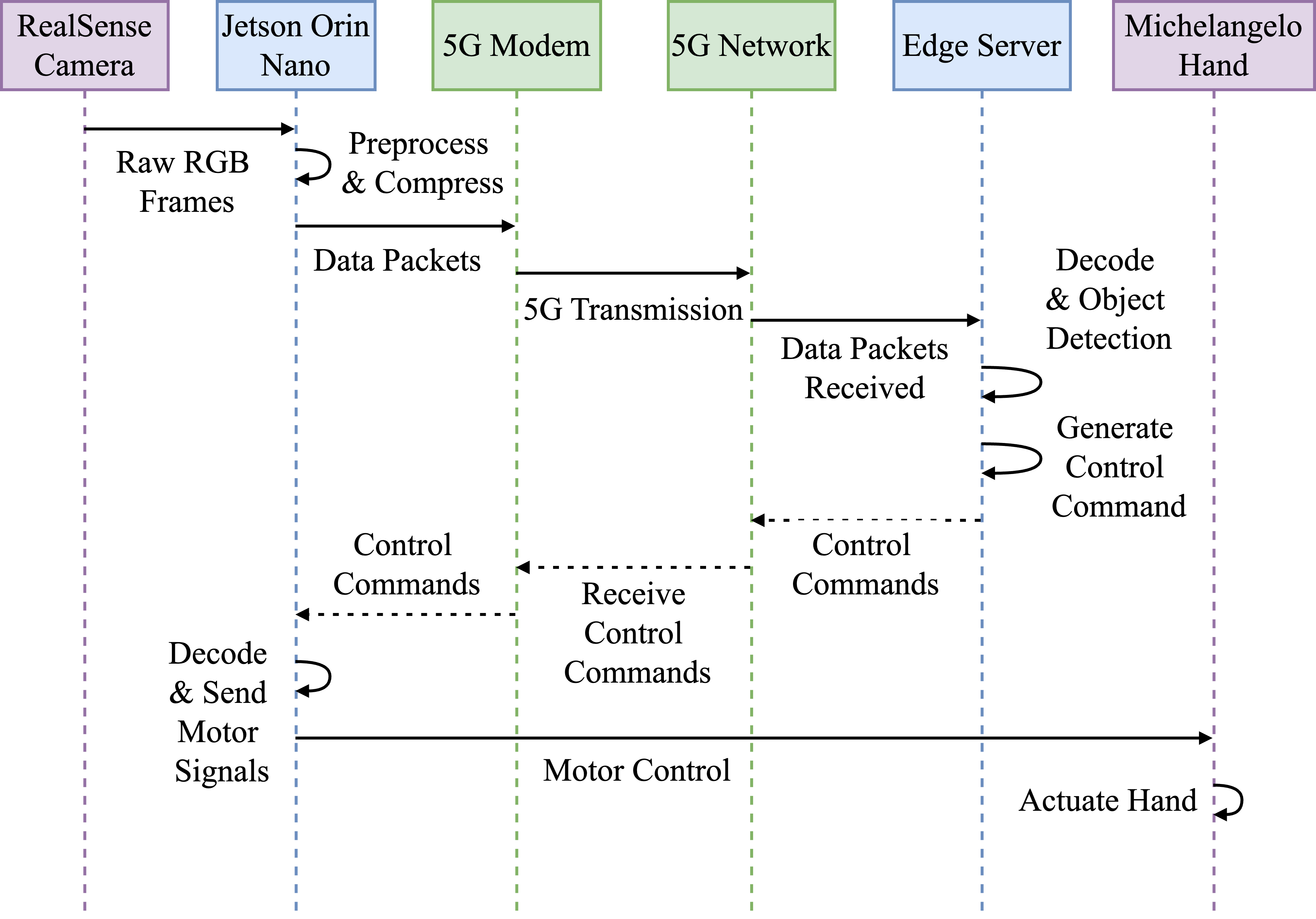} 
\caption{The processing pipeline.}
\label{fig:message_sequence_chart}
\end{figure}

The edge server features an AMD Ryzen 9 7950X 16-core processor, an NVIDIA Quadro P1000 GPU, 64 GB of RAM, and 1 TB of NVMe storage.
Besides the running of the 5G network, the server utilizes its GPU for real-time video processing and generates a control signal that represents the corresponding grasping pattern.
This is transmitted back to the prosthesis,
thereby closing the control loop.
An essential component of a smart prosthesis is an interface that allows the user to convey explicit (volitional) commands to the system, enabling semi-autonomous control. In our prototype, this is achieved using the Myoarm band from Thalmic Labs, a multichannel EMG system with a sampling rate of 200 Hz. As demonstrated in the literature \cite{markovic_sensor_2015}, myoelectric control is used to trigger automatic functions (e.g., performing a gesture to activate preshaping) or take over control (e.g., closing the hand manually after it has been automatically preshaped). The myoelectric interface was not used in the present study, which was focused on connectivity, as explained below. 

\subsection{Processing Pipeline and Control Flow}

The closed-loop control system follows the data processing pipeline depicted in Fig.~\ref{fig:message_sequence_chart} and elaborated as follows: 

\begin{enumerate}
    \item \textbf{Visual Data Acquisition:} The Intel RealSense D405 camera captures synchronized RGB and depth frames at a resolution of 640x480 pixels and a frame rate of 30 FPS, which represents a balance between visual fidelity and processing demands.
    \item \textbf{Local Preprocessing and 5G Transmission:} The Jetson device receives the raw video frames and performs initial preprocessing, where JPEG compression is applied to reduce payload size, thus optimizing the bandwidth usage before transmission.
    The compressed video frames are then transmitted to the edge server.
    \item \textbf{Object Detection at the Edge Server:}  Leveraging its 
    GPU unit, the edge server implements real-time object detection using the YOLOv8 model \cite{yolov8_ultralytics}. The model processes the received video frames and provides information about the scene captured by the prosthesis camera, such as the different objects and their spatial locations. 
    \item \textbf{Control Command Generation and Execution:} Based on the object detection results, the control logic, implemented at the edge server, generates the corresponding grip pattern (palmar or lateral).  The 5G base station transmits back the generated control signal to the local processing unit (i.e., the Jetson device), where it is decoded and translated into motor actuation signals for the desired grip pattern. 
\end{enumerate}

\begin{figure}[t]
\centering
\includegraphics[width=0.75\linewidth]{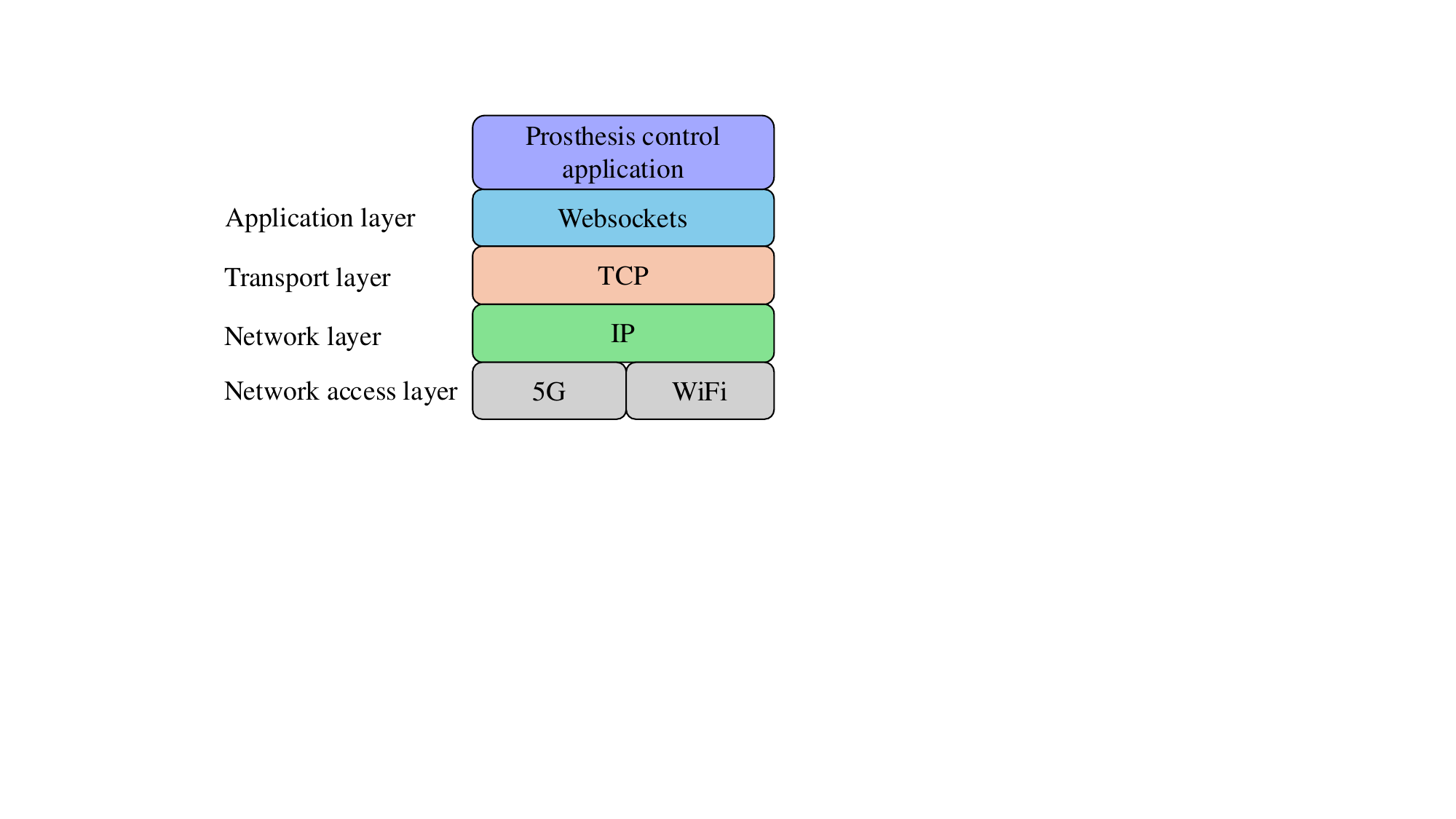}
\caption{The communication protocol stack.}
\label{CommStack}
\end{figure}
\subsection{Communication Protocol Stack}

The protocol stack employed in the system is shown in Fig.~\ref{CommStack}.
Specifically, the realized prototype employs WebSockets at the application layer, which run over TCP at the transport layer.\footnote{We note that, although UDP is often preferred over TCP for real-time media because of its reduced latency, we chose WebSockets/TCP solution to guarantee reliable communication 
that is advantageous for preserving data integrity, particularly in the initial development and testing stages.}
This solution was chosen as it offers a standardized framework for establishing persistent, bidirectional communication channels in client-server architecture. 
As already noted, the network access layer features two alternatives: 5G and WiFi. 

\subsection{Software Optimizations for Real-time Performance} \label{software_optimizations}

To achieve and maintain real-time performance in the system, particularly given the inherent latency of network communication, we implemented several key software optimizations in both the client and the server applications: 
\begin{itemize}
    \item \textbf{Asynchronous, Non-blocking Communication:}  The entire communication and processing pipeline is built using Python's asyncio library and the websockets library. This asynchronous approach enables non-blocking operations, allowing the client and server to concurrently send, receive, and process data without waiting for each operation to complete sequentially. %This is essential for pipelining and maximizing throughput.
    \item \textbf{Multi-Frame Pipelining:} 
    The system implements pipelining through asynchronous WebSockets communication, where frame transmission operates independently of frame reception. This means while frame $N$ is being transmitted to the server, the client can continue capturing and preparing frame $N+1$ while simultaneously receiving processed results for frame $N-1$. This concurrent operation is achieved through asyncio's gather mechanism, which allows multiple frames to be in different stages of processing simultaneously without blocking each other. 
    \item \textbf{JPEG Compression:} To minimize bandwidth consumption and reduce transmission latency, we utilize JPEG compression for video frames.  The client-side software adjusts the JPEG quality level to 90, which produces very high-quality images with a significant reduction in their size.
    \item \textbf{Server-Side Thread Pool Executor:}  On the edge server, the computationally intensive object detection task (i.e., YOLOv8 inference) is offloaded to a thread pool using ThreadPoolExecutor. This prevents the main asyncio event loop on the server from being blocked by long-running inference tasks, ensuring that the server remains responsive to incoming client requests and can efficiently handle concurrent frame processing.
\end{itemize}

\section{Evaluation}
In this section, we evaluate the connectivity performance of the proposed framework under two networking configurations: 5G (outdoor scenario) and WiFi (indoor scenario). The objective is to demonstrate the feasibility of the proposed framework to enable end-to-end real-time performance of a semi-automatically controlled prosthesis. 
%%%%%%%%%%%%%%%%%%%%%%%%%%%%%%%%%%
\subsection{The Experimental Setup}
%%%%%%%%%%%%%%%%%%%%%%%%%%%%%%%%%%
The conducted experiments involved transmitting JPEG-compressed video frames, captured by the Intel RealSense D405 camera, to the edge server, where they were processed for object detection and control command generation. Two data formats were considered for downlink transmissions: JSON, representing compact control commands of ca. 0.75 KB, and annotated video frames of ca. 60~KB representing visual feedback of the detected objects and their labels to be used in wearable augmented reality (AR) for a better user experience.
\begin{figure}[!t]
    \centering
    \includegraphics[width=1\linewidth]{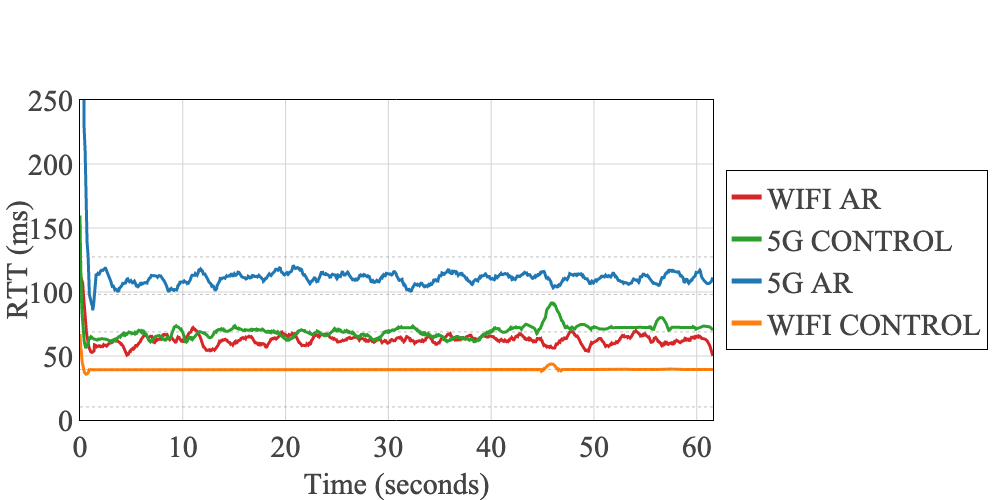}
    \caption{Round-Trip Time (RTT) under 5G and WiFi configurations.}
    \label{fig:rtt_over_time}
\end{figure}

%%%%%%%%%%%%%%%%%%%%%%%%%%%%%%%%%%
\subsection{Results}
%%%%%%%%%%%%%%%%%%%%%%%%%%%%%%%%%%
Fig.~\ref{fig:rtt_over_time} shows the obtained Round Trip Time (RTT) in our experimental setup for both 5G and WiFi configurations. The RTT is measured as the sum of the total link-level latency (uplink and downlink transmissions) and the processing time at the edge server. The RTT was obtained using timestamping at both the client side (prosthetic system) and the edge server. Note that the RTT excludes the pre-processing time at the local computing unit (NVIDIA Jetson Orin Nano), which is negligible ($\approx$ 2 ms) compared to the processing time at the edge server and the link-level latency. The results in Fig.~\ref{fig:rtt_over_time} illustrate that WiFi consistently achieves lower RTT than 5G for both the JSON-based control commands and the AR video frames. The slightly increased variability in 5G latency is mainly attributed to the increased link-level latency due to the fluctuations in the cellular radio channel conditions that are less pronounced in WiFi’s short-range, low-interference environment.

The average RTT and the corresponding standard deviation (STD) of both 5G and WiFi are given in Table~\ref{tab:performance_metrics_json_90} and Table~\ref{tab:performance_metrics_video_90} for the JSON-based control commands and AR video frames, respectively. The results demonstrate that the online control loop achieves an average RTT of 39.6 ms (STD = 2.2 ms) with WiFi connectivity, while it increases to 69.6 ms (STD = 11.1 ms) in the 5G setup. Despite the higher RTT in 5G, the end-to-end control latency remains within the 125 ms limit required for natural prosthetic control, validating the feasibility of the cloud-connected prosthetic framework. For the AR video frames, the average RTT increases to 62.5 ms (STD = 14.6 ms) and 112.2 ms (STD = 22.9 ms) for the WiFi and 5G, respectively. The increased RTT compared to the JSON-based control commands is due to the higher payload size of the video frames, which adds additional latency to the processing time at the edge server (compression of video frames) as well as the 5G/WiFi downlink.
The results listed in Tables~\ref{tab:performance_metrics_json_90} and~\ref{tab:performance_metrics_video_90} also report consistent server-side processing times for both 5G and WiFi configurations, which fulfill the 33 ms/frame budget for 30 FPS, emphasizing that the processing pipeline at the edge server does not introduce a potential bottleneck that would affect the real-time control.

\begin{table}[!t]
    \centering
    \caption{Average results for the JSON-based control commands}
    \label{tab:performance_metrics_json_90}
    \begin{tabular}{lcc}
        \hline
        \textbf{Metric} & \textbf{WiFi} & \textbf{5G} \\
        \hline
        End-to-end RTT (ms) & 39.57 $\pm$ 2.22 & 69.57 $\pm$ 11.14 \\
        Server Processing (ms) & 13.04 $\pm$ 0.40 & 12.13 $\pm$ 0.68 \\
        Message Rate (MPS) & 29.79 $\pm$ 2.01 & 29.84 $\pm$ 1.87 \\
        DL Bandwidth (Mbits/s) & 14.50 $\pm$ 0.99 & 14.64 $\pm$ 0.93 \\
        UL Bandwidth (Mbits/s) & 0.19 $\pm$ 0.02 & 0.18 $\pm$ 0.02 \\
        Frame Drop Rate & 3.47 $\pm$ 2.03 & 6.56 $\pm$ 3.22 \\
        \hline
    \end{tabular}
\end{table}

\begin{table}[!t]
    \centering
    \caption{Average results for the AR video streaming}
    \label{tab:performance_metrics_video_90}
    \begin{tabular}{lcc}
        \hline
        \textbf{Metric} & \textbf{WiFi} & \textbf{5G} \\
        \hline
        End-to-end RTT (ms) & 62.54 $\pm$ 14.59 & 112.18 $\pm$ 22.89 \\
        Server Processing (ms) & 12.98 $\pm$ 0.35 & 13.21 $\pm$ 0.37 \\
        Frame Rate (FPS) & 22.59 $\pm$ 1.58 & 23.24 $\pm$ 1.49 \\
        DL Bandwidth (Mbits/s) & 11.10 $\pm$ 0.78 & 11.33 $\pm$ 0.73 \\
        UL Bandwidth (Mbits/s) & 12.45 $\pm$ 0.96 & 12.69 $\pm$ 0.99 \\
        Frame Drop Rate & 5.62 $\pm$ 2.55 & 10.63 $\pm$ 4.06 \\
        \hline
    \end{tabular}
\end{table}

\begin{figure}[!t]
    \centering
    \includegraphics[width=1\linewidth]{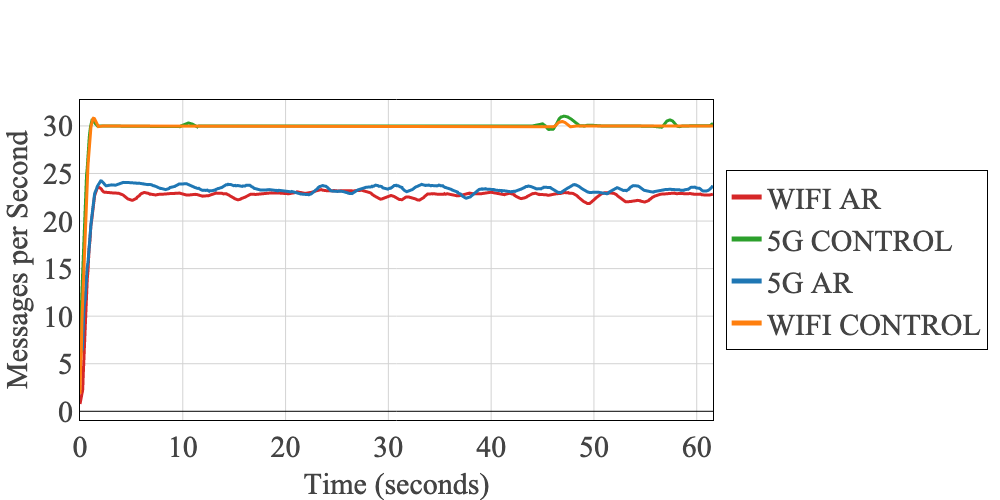}
    \caption{Frame/message rate over time under 5G and WiFi configurations.}
    \label{fig:fps_over_time}
\end{figure}

\begin{figure}[!t]
    \centering
    \includegraphics[width=1\linewidth]{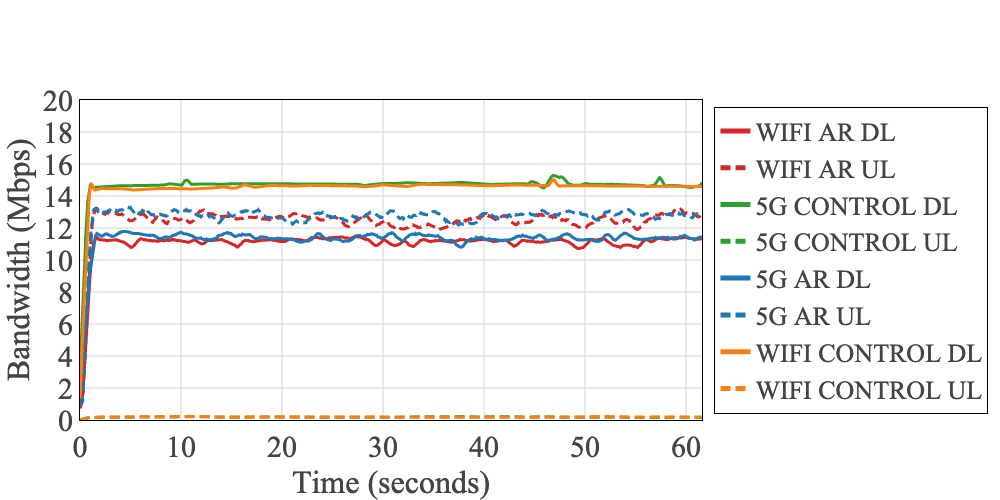}
    \caption{Bandwidth usage over time under 5G and WiFi configurations.}
    \label{fig:bandwidth_over_time}
\end{figure}

The frame/message rate serves as an indicator of the system’s ability to process and transmit data in real time. Fig.~\ref{fig:fps_over_time} presents the observed frame rates (FPS) and message rates (MPS) over time for both network configurations. The results indicate that JSON-based control commands maintain a stable message rate of approximately 29.8 MPS across both 5G and WiFi, ensuring smooth and uninterrupted control signal transmission. In the case of AR video streaming, the frame rate was 22.59 FPS over WiFi and 23.24 FPS over 5G, with 5G exhibiting slightly higher fluctuations due to network variability. The difference in frame rates between JSON and AR video transmission is expected, as AR video processing involves additional computational overhead at the edge server, particularly for object detection and annotation. Frame drop rates were slightly higher in 5G than WiFi, with 10.6\% of AR video frames dropped over 5G, compared to 5.6\% over WiFi. For JSON-based control commands, the frame drop rate was 6.6\% over 5G and 3.5\% over WiFi, indicating greater stability in WiFi’s controlled environment. The same trend is also observed for the uplink and downlink bandwidth usage as depicted by Fig.~\ref{fig:bandwidth_over_time}. For the JSON control commands, both 5G and WiFi achieve approximately 0.2 Mbits/s and 14.5 Mbits/s for the uplink and downlink bandwidth, respectively, while they achieve 12.5 Mbits/s uplink and 11 Mbits/s downlink for AR video frames. 

\begin{figure}[!t]
\centering
\begin{subfigure}[t]{0.32\linewidth}
        \centering
\includegraphics[width=\linewidth]{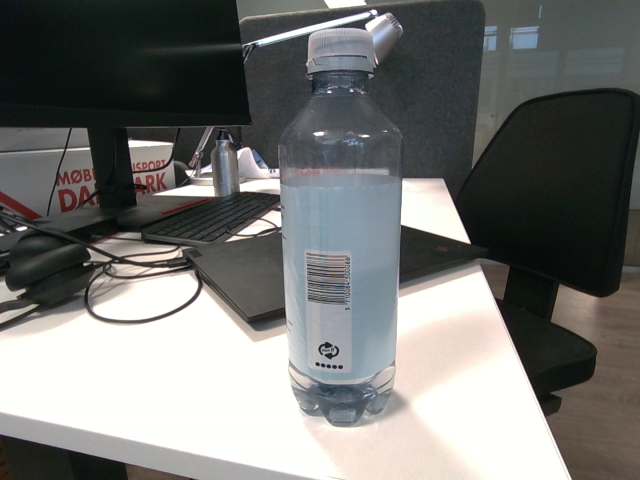}
\caption{Original image}
\end{subfigure}
 \hfill
\begin{subfigure}[t]{0.32\linewidth}
 \centering
\includegraphics[width=\linewidth]{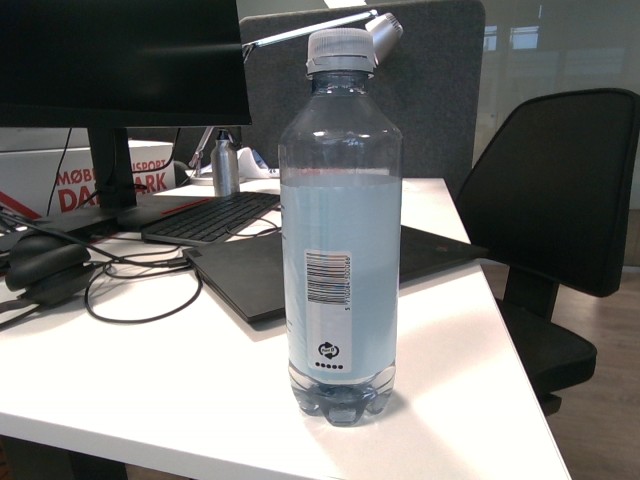}
\caption{Compression at the client-side}
\end{subfigure}
\hfill
\begin{subfigure}[t]{0.32\linewidth}
\includegraphics[width=\linewidth]{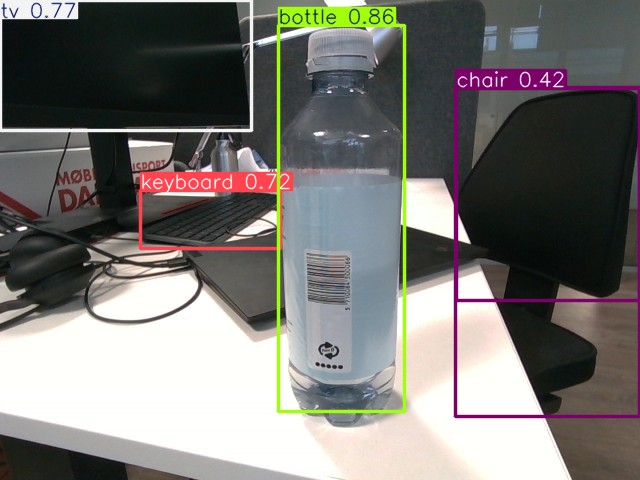}
\caption{Compression at the server-side}
\end{subfigure}
\caption{Image quality under the adopted JPEG compression with Q90.}
\label{fig:compression_stages}
\end{figure}

JPEG compression is used in our system to reduce bandwidth requirements for data transmission. At both the prosthetic system (client-side) and edge server (server-side), in case of sending back the AV video frames, we used JPEG compression with a quality factor of 90, striking a balance between efficient data transfer and maintaining sufficient image quality for reliable object detection. Fig.~\ref{fig:compression_stages} showcases the impact of dual-stage compression at different processing stages. The original raw camera capture of 364~KB per image was compressed to~50 KB at the client-side before transmission. After edge-side processing, where bounding boxes and labels were added for AR visualization, the frame size increased slightly to 60 KB. Despite the significant reduction in data size, the image quality remained intact, ensuring clear object recognition and accurate user feedback. 

\section{Discussion}

In this paper, we described a general framework that can be used for the development and testing of cloud and edge-connected prosthetic systems.
The framework includes a prosthesis equipped with a computer vision sensor and an edge device implementing local processing and communication with the edge server for computational offloading. The setup was used to assess the feasibility of establishing online control loop where the data were acquired from the prosthesis, pre-processed, and sent to the server, which computed commands (JSON) and feedback (video), and transmitted them back to the hand. The aim was to test if the overall round-trip time was below the threshold for smooth prosthesis response reported in the literature. The test considered data flow and algorithms representative of a typical pipeline for semi-automatic prosthesis control \cite{ghazaei_deep_2017}, in which video frames were analyzed using a common machine-learning approach for object detection and recognition (YOLO). 

The results show that the average end-to-end latency was indeed consistently below 125 ms when using both link types (WiFi and direct 5G connection). This is an important outcome, demonstrating that the concept of cloud-connected prostheses (see, e.g., \cite{chiariotti_future_2024}) can be implemented both indoors (WiFi) and outdoors (5G) with round-trip latencies well below the critical threshold. 
In addition, the low server-side processing times across all configurations validate the efficiency of the processing pipeline.
This means that the computationally intensive processing can indeed be offloaded to the edge/cloud without negatively affecting the speed of the prosthesis response. Therefore, the local control loop can be "rerouted" to the edge/cloud transparently to the user and without decreasing the quality of their experience when interacting with the device. This is an important and encouraging result for the further development and implementation of cloud-connected prosthetic systems.

In practical use, the performance of the system is fundamentally connected to network availability and reliability, especially in relation to 5G coverage. The evaluation conducted under controlled network conditions revealed superior latency and throughput; nonetheless, actual 5G network performance may show variability. 
While extensive testing in various real-world environments 
% (e.g., crowded urban areas, indoor-outdoor transitions) 
was beyond the scope of this proof-of-concept study, our evaluation provides the foundation for understanding the system's performance boundaries. % if we want to answer to reviewer id 5 (3rd)
To address this in future practical applications, a failsafe mechanism has to be implemented to reduce network dependency, leveraging local processing capabilities to maintain essential prosthetic functions during network interruptions. Investigating predictive network quality monitoring and hybrid methodologies that adaptively alternate between cloud-based and local control in response to network conditions is part of our future work.

In addition, a practical edge/cloud-connected prosthetic system requires the use of privacy and security-preserving mechanisms, as well as the characterization and potential optimization of power consumption at the user's side to maintain a satisfactory level of device autonomy.
Investigations of these aspects are also left for future work.

It is important to mention that EMG signals, managed by the local controller, were excluded from latency evaluation as this study focuses on cloud-based vision processing.
Nevertheless, the presented results suggest that there is a substantial latency budget remaining for the accommodation of the execution of this task.
In addition, Future cloud implementations with EMG processing will require QoS and network slicing to prioritize control signals.
 % answer to reviewer id 3 (2nd)

Finally, we note that the next step in this research is the implementation of a fully operational prosthesis prototype that can be used to perform functional tasks. This will allow for assessing the impact of edge/cloud-connected prostheses on the device utility and user experience. 
\section*{Acknowledgement}

The work presented in this paper was supported by the Independent Research Fund Denmark (DFF), project no. 2035-00169B ``CLIMB''.
This research did not involve any human subjects or animal experiments.

\bibliographystyle{IEEEtran}
\bibliography{references}

\end{document}